\documentclass[twocolumn,showpacs,amsmath,amssymb,bibnotes]{revtex4}
\usepackage{bm}
\usepackage{subfigure}
\usepackage[dvips]{graphics,epsfig}

\newcommand{\be}{\begin{equation}}
\newcommand{\ee}{\end{equation}}
\newcommand{\bea}{\begin{eqnarray}}
\newcommand{\eea}{\end{eqnarray}}
\newcommand{\gdot}{\dot{\gamma}}
\newcommand{\gdotbar}{\overline{\dot{\gamma}}}

\newcommand{\bw}{\begin{widetext}}
\newcommand{\ew}{\end{widetext}}

\newcommand{\Gmodulus}{G_0}

\newcommand{\vecv}[1]{\mathbf{{#1}}}
\newcommand{\tens}[1]{\mathbf{{#1}}}
\newcommand{\nablu}{{\bf \nabla}}

\begin{document}

\title{Shear banding and interfacial instability in planar Poiseuille flow} 
\author{Suzanne M. Fielding$^1$
  and Helen J. Wilson$^2$}
\email{suzanne.fielding@durham.ac.uk, helen.wilson@ucl.ac.uk}
\affiliation{$^1$ Department of Physics, University of Durham, Science Laboratories, South Road, Durham, DH1 3LE, United Kingdom\\ $^2$
  Department of Mathematics, University College London, Gower Street,
  London WC1E 6BT, United Kingdom }

\date{\today}
\begin{abstract}
  Motivated by the need for a theoretical study in a planar geometry
  that can easily be implemented experimentally, we study the pressure
  driven Poiseuille flow of a shear banding fluid. After discussing
  the ``basic states'' predicted by a one dimensional calculation that
  assumes a flat interface between the bands, we proceed to
  demonstrate such an interface to be unstable with respect to the
  growth of undulations along it. We give results for the growth rate
  and wavevector of the most unstable mode that grows initially, as
  well as for the ultimate flow patterns to which the instability
  leads. We discuss the relevance of our predictions to the present
  state of the experimental literature concerning interfacial
  instabilities of shear banded flows, in both conventional rheometers
  and microfluidic channels.

\end{abstract}
\pacs{{47.50.+d}, %{ Non-Newtonian fluid flows}--
     {47.20.-k}, %{ Hydrodynamic stability}--
     {36.20.-r}.%{ Macromolecules and polymer molecules}
     } 
\maketitle

%%%%%%%%%%%%%%%%%%%%%%%%%%%%%%%%%%%%%%%%%%%%%%%%%%%%%%%%%%%%%%%%%%%%%%%%%%%%%

\section{Introduction}
\label{sec:intro}

Complex fluids have internal mesoscopic structure that is readily
reorganised by an imposed shear flow. This reorganisation in turn
feeds back on the flow field, resulting in strongly nonlinear
constitutive properties. In some systems this nonlinearity is so
pronounced that the underlying constitutive curve relating shear
stress $T_{xy}$ to shear rate $\gdot$ in homogeneous flow is predicted
to have a region of negative slope
$dT_{xy}/d\gdot<0$~\cite{spenley93,spenley96}. In this regime, an
initially homogeneous flow is unstable to the formation of coexisting
shear bands of differing local viscosities and internal structuring,
with band normals in the flow-gradient direction
$y$~\cite{Yerushalmi70}.  The signature of this transition in bulk
rheometry is the presence of characteristic kinks, plateaus and
non-monotonicities in the composite flow
curve~\cite{rehage91}. Explicit observation of the bands is made using
local rheological techniques such as flow birefringence~\cite{Decr+95}
and NMR~\cite{britton-prl-78-4930-1997,callaghan2008},
ultrasound~\cite{BecManCol04,manneville2008}, heterodyne dynamic light
scattering~\cite{SalColManMol03,manneville2008} or particle
image~\cite{hu-jr-49-1001-2005} velocimetry. Using these methods, the
existence of shear banding has been firmly established in a wide range
of complex fluids, including
wormlike~\cite{BRP94,callaghan96,grand97,britton-prl-78-4930-1997,berret94a,schmitt94,Capp+97,rehage91,Decr+95,Makh+95,BPD97}
and
lamellar~\cite{diat93,wilkinsOlms2006,Bonn+98,SalManCol03,SalManCol03b,ManSalCol04}
surfactants; side-chain liquid crystalline polymers~\cite{pujolle01};
viral
suspensions~\cite{lettinga-jpm-16-S3929-2004,dhont-jcp-118-1466-2003};
telechelic polymers~\cite{berret-prl-8704--2001}; soft
glasses~\cite{CRBMGHJL02,holmes-jr-48-1085-2004,rogers2008}; polymer
solutions~\cite{HilVla02}; and colloidal
suspensions~\cite{ChenZAHSBG92}.

Beyond the basic observation of shear banding, experiments with
enhanced spatial and temporal resolution have more recently revealed
the presence of complex spatio-temporal patterns and dynamics in many
shear banded
flows~\cite{BecManCol04,holmes-el-64-274-2003,lopez-gonzalez-prl-93--2004,bandyopadhyay-prl-84-2022-2000,ganapathy-prl-96--2006,HBP98,bandyopadhyay-el-56-447-2001,wheeler-jnfm-75-193-1998,herle-l-21-9051-2005,fischer-ra-39-234-2000,azzouzi-epje-17-507-2005,decruppe-pre-73--2006,SalManCol03b,Salmon02,ManSalCol04,HilVla02,lerouge-prl-96--2006,lerouge2008,fardin2009}.
In many such cases, the bulk stress response of the system to a steady
imposed shear rate (or vice versa) is intrinsically unsteady, showing
either temporal oscillations or erratic fluctuations about the average
(flow curve) value.  Local rheological measurements reveal such
signals commonly to be associated with a complicated behaviour of the
interface between the
bands~\cite{BecManCol04,holmes-el-64-274-2003,lopez-gonzalez-prl-93--2004,HBP98,wheeler-jnfm-75-193-1998,herle-l-21-9051-2005,fischer-ra-39-234-2000,SalManCol03b,HilVla02,lerouge-prl-96--2006,lerouge2008,fardin2009}.
The majority of these measurements have been in one spatial dimension
(1D), normal to the interface between the bands. However 2D
observations in Refs.~\cite{lerouge-prl-96--2006,lerouge2008}
explicitly revealed the presence of undulations along the interface,
in a boundary driven curved Couette flow, accompanied by Taylor-like
vortices~\cite{fardin2009}. The undulations were shown to be either
static or dynamic, according to the imposed flow parameters.

Theoretically, instability of an initially flat interface between
shear bands was predicted in boundary driven planar Couette flow in
Refs.~\cite{fielding-prl-95--2005,wilson-jnfm-138-181-2006,fielding2007b}. In
this work, separate 2D studies in the flow/flow-gradient
($x$-$y$)~\cite{fielding-prl-95--2005,wilson-jnfm-138-181-2006} and
flow-gradient/vorticity ($y$-$z$)~\cite{fielding2007b} planes revealed
instability with respect to undulations along the interface with
wavevector in the flow and vorticity directions respectively.  In both
cases the mechanism for instability was suggested to be a jump in
normal stress across the interface~\cite{hinch-jnfm-43-311-1992}.

While these predictions provide a good starting point, there remains
the possibility that the interfacial undulations observed in
Refs.~\cite{lerouge-prl-96--2006,lerouge2008,fardin2009} originate
instead in curvature driven effects such as a bulk viscoelastic
instability of the Taylor Couette~\cite{larson-jfm-218-573-1990} type
in the high shear band, as discussed in Ref.~\cite{fardin2009}. These
were neglected in the planar calculations of
Refs.~\cite{fielding-prl-95--2005,wilson-jnfm-138-181-2006,fielding2007b}
(Other possibilities, also neglected, include free surface
instabilities at the open rheometer edges; and an erratic stick-slip
motion at the solid walls of the flow cell. We shall not consider
these further in what follows here either.)

In principle, therefore, either (or both) of (at least) two possible
mechanisms could underlie the observed interfacial undulations: (i) a
bulk viscoelastic Taylor Couette like instability of the strongly
sheared band, or (ii) instability of the interface between the bands,
driven by the normal stress jump across it. Of these, scenario (i) can
only arise in a curved geometry.  Experiments in planar shear could
therefore in principle help discriminate between these scenarios, by
eliminating the curvature required for (ii). However they are
technically difficult to implement in a boundary driven setup.

There thus exists a clear need for theoretical predictions in a planar
flow geometry that could easily be implemented experimentally. An
obvious candidate comprises pressure driven flow in a rectilinear
microchannel of rectangular cross section with a high aspect ratio
$L_z/L_y\gg 1$. Indeed, such experiments have recently been
performed~\cite{nghe2008,masselon2008,submitted}, as discussed in more
detail below. With this motivation in mind, in this paper we study the
planar Poiseuille flow of a shear banding fluid driven along the main
flow direction $x$ by a constant pressure drop $\partial_xP=-G$. For
simplicity we assume the fluid to be sandwiched between stationary
infinite parallel plates at $y=\{0,L_y\}$, neglecting the lateral
walls in the $z$ direction, and so taking the limit $L_z/L_y\to\infty$
at the outset. Our main contribution will be to show an interface
between shear bands to be unstable in this pressure driven geometry,
as it is in the boundary driven planar Couette flow studied
previously~\cite{fielding-prl-95--2005,wilson-jnfm-138-181-2006,fielding2007b}. We
will furthermore give results for the growth rate and wavevector
associated with the early stage kinetics of this instability, as well
as the ultimate flow patterns to which it leads.

\begin{figure}[tb!]
  \includegraphics[width=8.0cm]{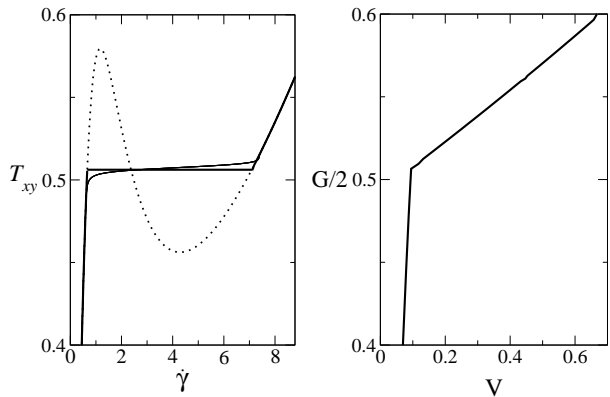}
  \caption{{\bf Left)} Dotted line: homogeneous constitutive curve for
    $a=0.3$, $\eta=0.05$. Thick solid line: composite flow curve for
    one-dimensional planar shear banded Couette flow (data already
    published in Ref.~\cite{fielding-prl-95--2005}). Selected stress
    $T_{\rm sel}=0.506$.  Thin solid line: parametric plot of local
    shear stress
    $T_{xy}(y)=\Sigma_{xy}(y)+\eta\gdot(y)=G(\tfrac{1}{2}-y)$ against
    local shear rate $\gdot(y)$ for one-dimensional planar shear
    banded Poiseuille flow with $l=0.00125$, $G=2.0$.\\ {\bf Right)}
    Halved pressure gradient versus total throughput for
    one-dimensional planar Poiseuille flow with $a=0.3$, $\eta=0.05$,
    $l=0.0025$. This shows a kink at the onset of banding at
    $G/2=T_{\rm sel}$ as expected.}
% thin solid line on left: base/raw/PPF_a0.3_eta0.05_l0.00125_G2.0_Npoints1_tmax500.0_dt0.01_ny3200
% throughput curve on right: base/rawGsweep_a0.3_eta0.05_l0.0025_Gmin0.1_Gmax10.0_Npoints40_tmax100.0_dt0.01_Ny1600.dat,Gsweep_a0.3_eta0.05_l0.0025_Gmin0.1_Gmax10.0_Npoints40_tmax100.0_dt0.01_Ny800.dat,Gsweep_a0.3_eta0.05_l0.0025_Gmin0.95_Gmax1.2_Npoints40_tmax100.0_dt0.01_Ny800.dat,base/raw/Gsweep_a0.3_eta0.05_l0.005_Gmin0.1_Gmax10.0_Npoints40_tmax100.0_dt0.01_Ny800.dat
\label{fig:flowCurve}
\vspace{-0.5cm}
\end{figure}

The paper is structured as follows. After introducing the rheological
model and boundary conditions in Sec.~\ref{sec:model}, we calculate in
Sec.~\ref{sec:base} the one-dimensional (1D) shear banded states that
are predicted when spatial variations are permitted only in the flow
gradient direction $y$, artificially assuming translational invariance
in $x$ and $z$, and accordingly assuming a flat interface between the
bands. These form the ``basic states'' and initial conditions to be
used in the stability calculations of the rest of the paper.

In Sec.~\ref{sec:xy} we study the linear stability of these 1D basic
states with respect to small amplitude perturbations with wavevector
$q_x\hat{\vecv{x}}$ in the flow direction. As in the case of boundary
driven flow studied previously, we find an undulatory instability of
the interface between the
bands~\cite{fielding-prl-95--2005,wilson-jnfm-138-181-2006,fielding2007b}. Results
are then presented for the ultimate nonlinear dynamical attractor in
this $x$-$y$ plane, from simulations that adopt periodic boundaries in
$x$. This exhibits interfacial undulations of finite amplitude that
convect along the flow direction at a constant speed.  In
Sec.~\ref{sec:yz} we turn instead to the flow-gradient/vorticity plane
$y$-$z$, likewise demonstrating linear instability of the interface
with respect to small amplitude perturbations with wavevector
$q_z\hat{\vecv{z}}$. We also give results for the ultimate nonlinear
flow state, which in this plane is steady. Directions for future work,
which will include full 3D calculations, are discussed in
Sec.~\ref{sec:conclusion}.

\section{Model and geometry}
\label{sec:model}

The generalised Navier--Stokes equation for a viscoelastic material in
a Newtonian solvent of viscosity $\eta$ and density $\rho$ is
\begin{equation} 
\label{eqn:NS} 
\rho(\partial_t +
  \vecv{v}.\nablu)\vecv{v} = \nablu.(\tens T -P\tens{I}) = \nablu .(\tens{\Sigma} + 2\eta\vecv{D}
  -P\tens{I}),
\end{equation} 
where $\vecv{v}(\vecv{r},t)$ is the velocity field and $\tens{D}$ is
the symmetric part of the velocity gradient tensor, $(\nablu
\vecv{v})_{\alpha\beta}\equiv \partial_\alpha v_\beta$. Throughout we
will assume zero Reynolds' number $\rho=0$. The pressure field
$P(\vecv{r},t)$ is determined by enforcing incompressibility,
\be
\label{eqn:incomp}
\vecv{\nabla}\cdot\vecv{v}=0.
\ee
The quantity $\vecv{\Sigma}(\vecv{r},t)$ in Eqn.~\ref{eqn:NS} is the
extra stress contributed to the total stress $\vecv{T}(\vecv{r},t)$ by
the viscoelastic component. We assume this to obey the diffusive
Johnson-Segalman (DJS)
model~\cite{johnson-jnfm-2-255-1977,olmsted-jr-44-257-2000}
\begin{widetext}
\be
\label{eqn:DJS}
(\partial_t
+\vecv{v}\cdot\nablu )\,\tens{\Sigma} 
= a(\tens{D}\cdot\tens{\Sigma}+\tens{\Sigma}\cdot\tens{D}) +
(\tens{\Sigma}\cdot\tens{\Omega} + \tens{\Omega}\cdot\tens{\Sigma})  
 + 2 \Gmodulus\tens{D}-\frac{\tens{\Sigma}}{\tau}+ \frac{\ell^2}{\tau }\nablu^2 
 \tens{\Sigma}.
\ee
\end{widetext}
Here $a$ is a slip parameter, $\Gmodulus$ is a plateau modulus, $\tau$
is the viscoelastic relaxation time, and $\tens{\Omega}$ is the
antisymmetric part of the velocity gradient tensor. The diffusive term
$\nablu^2 \tens{\Sigma}$ is needed to correctly capture the structure
of the interface between the shear bands, with a slightly diffusive
interfacial thickness $O(l)$, and to ensure unique selection of the
shear stress at which banding occurs~\cite{lu-prl-84-642-2000}.

Within this model we study flow between infinite flat parallel plates
at $y=\{0,L_y\}$. The fluid is driven in the positive $x$ direction by
a constant pressure gradient $\partial_x p = -G$, the plates being
held stationary.  At the plates we assume conditions of zero flux
normal to the wall
$\hat{\vec{n}}.\nablu\Sigma_{\alpha\beta}=0\;\forall\;\alpha,\beta$
for the viscoelastic stress (although other choices are
possible~\cite{adams2008}), with no slip and no permeation for the
fluid velocity. Throughout we use units in which $\Gmodulus=1,\tau=1$
and $L_y=1$. Unless otherwise stated we use a (dimensionless) value
$\eta=0.05$, suggested by the experiments of Refs.~\cite{nghe2008}. We
take $a=0.3$, although our results are quite robust with respect to
variations in this quantity. An order of magnitude estimate suggests
$l=O(10^{-3})$~\cite{fielding-epje-11-65-2003}.

\section{One-dimensional basic state}
\label{sec:base}

In this section we discuss the flow curves and shear banded states
predicted by 1D calculations that allow spatial variations only in the
flow-gradient direction $y$, assuming (often artificially)
translational invariance in the flow direction $x$ and vorticity
direction $z$. As a warm-up discussion to the case of planar
Poiseuille flow (PPF) that forms the primary interest of this paper,
we review first the more familiar case of planar Couette flow (PCF).

In steady 1D PCF, the shear stress $T_{xy}$ is uniform across the flow
cell. This follows trivially from solving Eqn.~\ref{eqn:NS} in this
geometry.  Within the (often artificial) assumption of a similarly
homogeneous shear rate field, corresponding to a velocity field
$\vecv{v}=\gdot y \hat{\vec{x}}$, the constitutive relation is then
given by $T_{xy}(\gdot)=\Sigma_{xy}(\gdot)+\eta\gdot$, where
$\Sigma_{xy}(\gdot)$ follows as the $xy$ component of the solution of
Eqn.~\ref{eqn:DJS} obtained by assuming stationarity in time and
homogeneity in space. See the dotted line in Fig.~\ref{fig:flowCurve}
(left). For an applied shear rate in the region of negative slope
$dT_{xy}/d\gdot<0$, homogeneous flow is predicted to be linearly
unstable with respect to the formation of shear bands. The steady
state composite flow curve is then shown by the thick solid line in
the same figure. For shear stresses $T_{xy}<T_{\rm sel}$
(resp. $T_{xy}>T_{\rm sel}$), where the selected stress $T_{\rm
  sel}=0.506$ for the particular choice of parameters in
Fig.~\ref{fig:flowCurve}, the system shows homogeneous flow on the low
shear (resp. high shear) branch of the constitutive curve.  Applying a
value of the shear rate $\gdotbar$ in the window between these
branches, we obtain shear bands that coexist at the selected stress
$T_{xy}=T_{\rm sel}$.

\begin{figure}[t]
  \includegraphics[width=8.0cm]{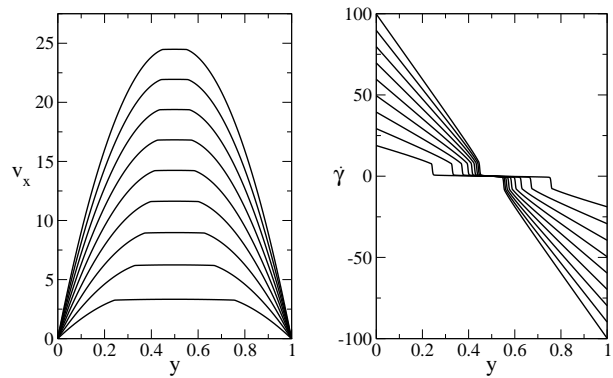}
  \caption{{\bf Left)} Velocity profiles in one-dimensional planar
    shear banded Poiseuille flow for $a=0.3$, $\eta=0.05$,
    $l=0.0025$. Increased throughputs correspond to increased applied
    pressure gradients $G=2.0$, 3.0, 4.0, 5.0, 6.0, 7.0, 8.0, 9.0,
    10.0.\\ {\bf Right)} Corresponding shear rate profiles, which follow as
    the spatial derivative of the velocity profiles.}
% base/raw/PPF_a0.3_eta0.05_l0.0025_G{2,3,4,5,6,7,8,9,10}.0_Npoints1_tmax50.0_dt0.001_ny800 
\label{fig:base}
\end{figure}

\begin{figure}[tb]
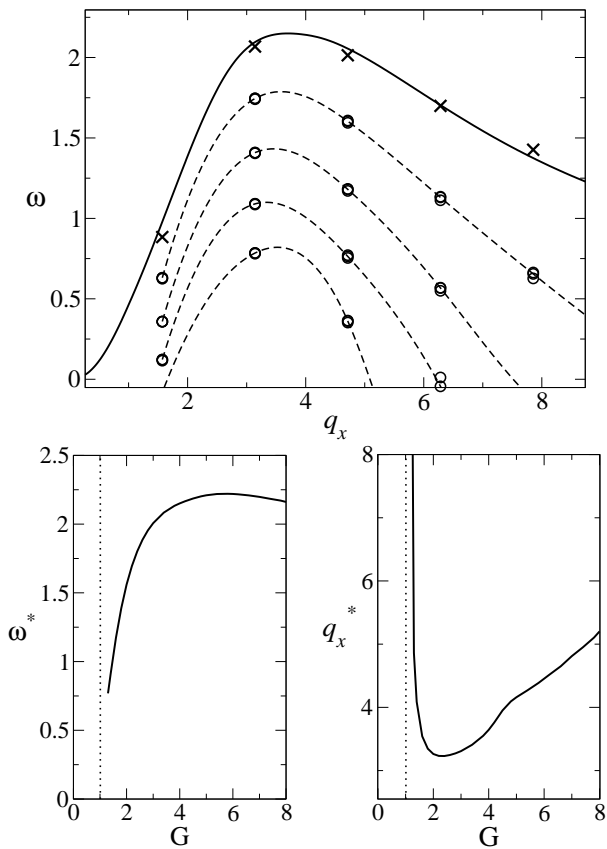

  \includegraphics[width=7.5cm]{./dispersion.eps}
  \includegraphics[width=8cm]{./peak.eps}
  \caption{{\bf Top)} Dispersion relation of growth rate $\omega$
    versus wavevector $q_x$ during the initial stage of 2D
    instability, starting with a 1D shear banded planar Poiseuille
    basic state. In each case $a=0.3$, $\eta=0.05$, $G=4.0$. Circles:
    extracted from early time dynamics of the full nonlinear
    code. Groups of data upward correspond to $l=0.02$, 0.015, 0.01,
    0.005.  Within each group, data are shown for $(Dt, N_x,
    N_y)=(0.000025, 200, 400)$, (0.0000125, 200, 400), (0.000025, 400,
    800).  These are mostly indistinguishable, demonstrating
    convergence with respect to grid and timestep. Dashed lines: cubic
    splines through the data for $(Dt, N_x, N_y)=(0.000025, 200,
    400)$, as a guide to the eye.  $L_x=4.0$ in each case. Solid line:
    results of analytical calculation in the true limit $l\to
    0$. Crosses: extrapolation of the $l\neq 0$ results to $l=0$ using
    the scaling $\omega(q_x,l)=\omega(q_x,l=0)-a(q_x)l$.  \newline
    {\bf Bottom)} Growth rate (left) and wavevector (right) at the peak
    of dispersion relations as a function of the pressure drop $G$,
    calculated analytically in the limit $l\to 0$.}
\label{fig:dispersion}
\end{figure}

\begin{figure}[tb]
  \includegraphics[scale=0.15]{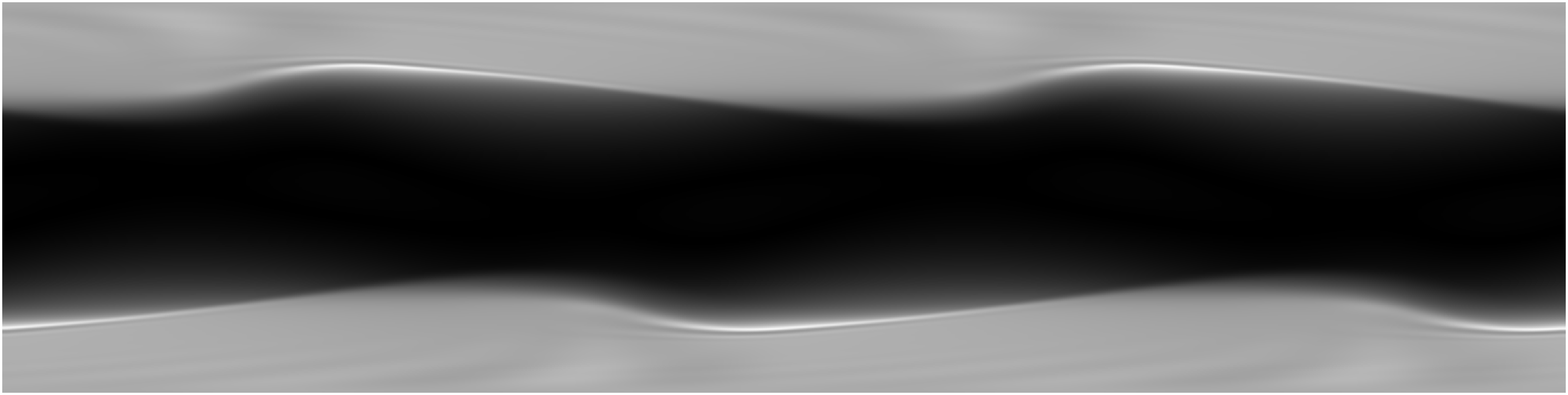}
  \includegraphics[scale=0.15]{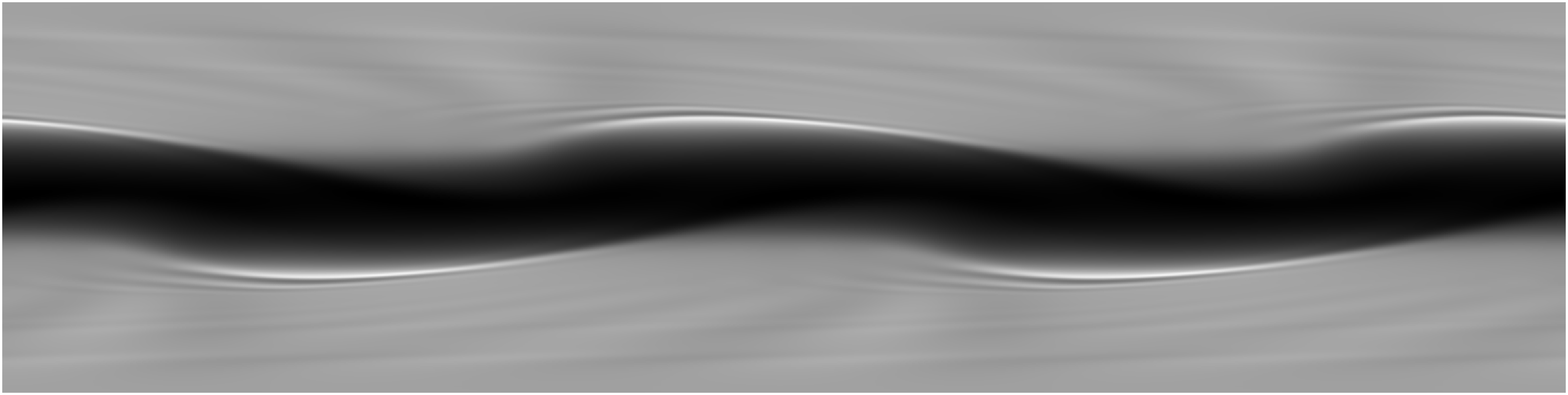}
  \includegraphics[scale=0.15]{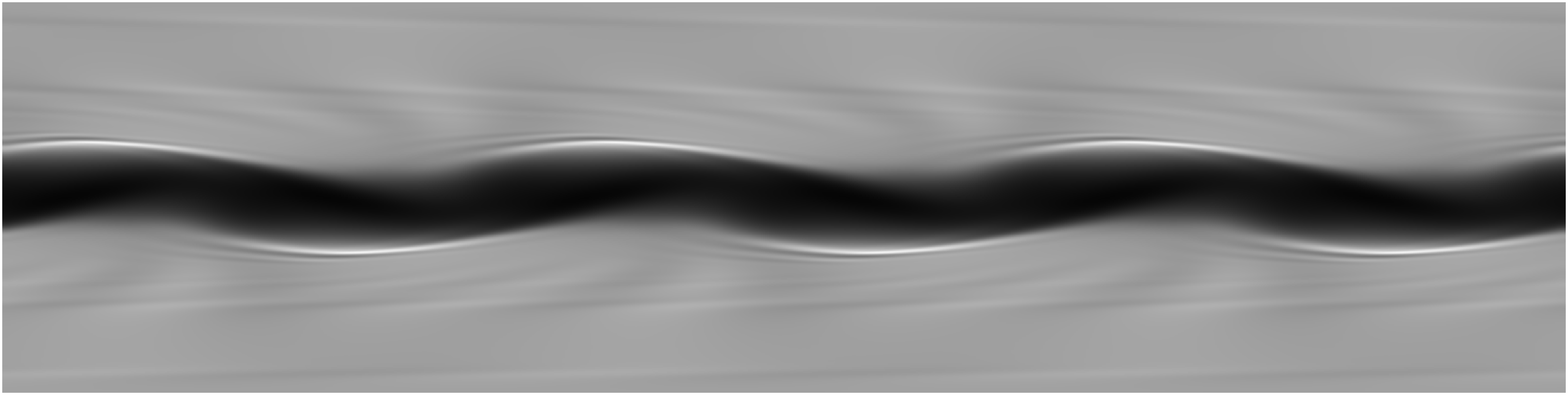}
  \includegraphics[scale=0.15]{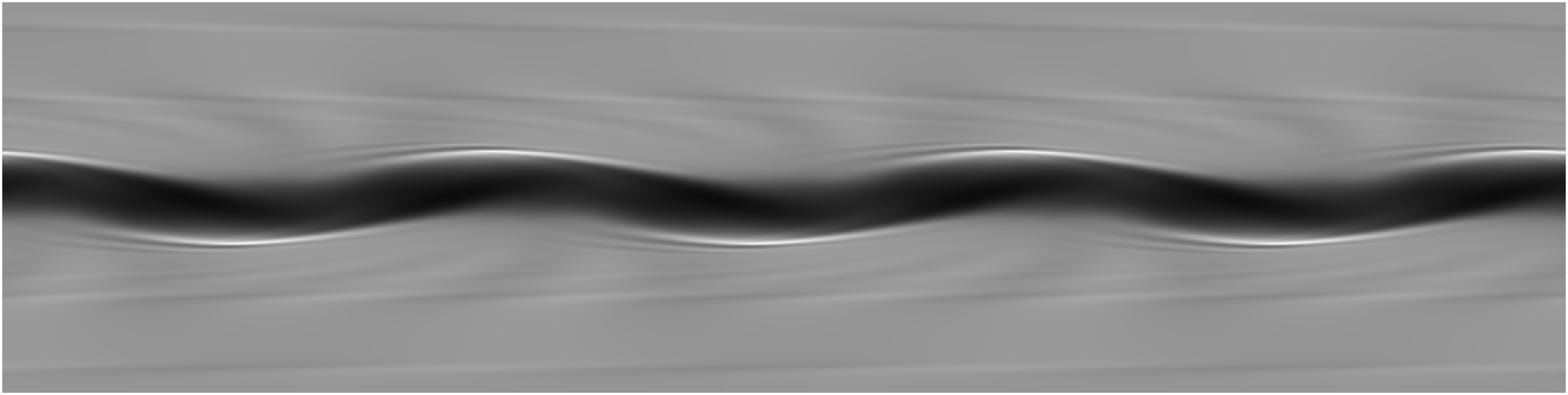}
  \includegraphics[scale=0.1]{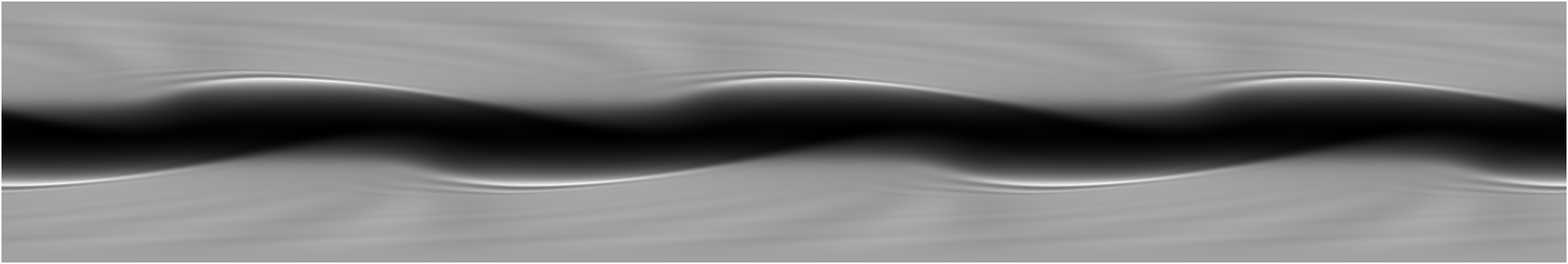}
  \caption{Greyscale snapshots of the order parameter
    $\Sigma_{xx}(x,y)$ at a representative time $t=60.0$ on the
    ultimate nonlinear attractor for $a=0.3$, $\eta=0.05$, $l=0.005$
    for $L_x=4.0$ and $G=2.0$, 4.0, 6.0, 8.0 (top four subfigures
    downwards) with $Dt=0.000025$, $N_x=200$, $N_y=400$. Bottom
    subfigure is for $G=4.0$ with a longer cell $L_x=6.0$ (so
    $N_x=300$).}
% statePlots/state_G2.0_gdot0.0_eta0.05_a0.3_l0.005_Lx4.0_temperature1e-14_Dt0.000025_Nx200_Ny400_tmax100.0_faster_time60.000000001711_OPWxx.eps
% statePlots/state_G4.0_gdot0.0_eta0.05_a0.3_l0.005_Lx4.0_temperature1e-14_Dt0.000025_Nx200_Ny400_tmax100.0_faster_time60.000000001711_OPWxx.eps
% statePlots/state_G6.0_gdot0.0_eta0.05_a0.3_l0.005_Lx4.0_temperature1e-14_Dt0.000025_Nx200_Ny400_tmax100.0_faster_time60.000000001711_OPWxx.eps
% statePlots/state_G4.0_gdot0.0_eta0.05_a0.3_l0.005_Lx6.0_temperature1e-14_Dt0.000025_Nx300_Ny400_tmax100.0_time60.000000001711_OPWxx.eps
\label{fig:snapshots}
\end{figure}

\begin{figure}[t]
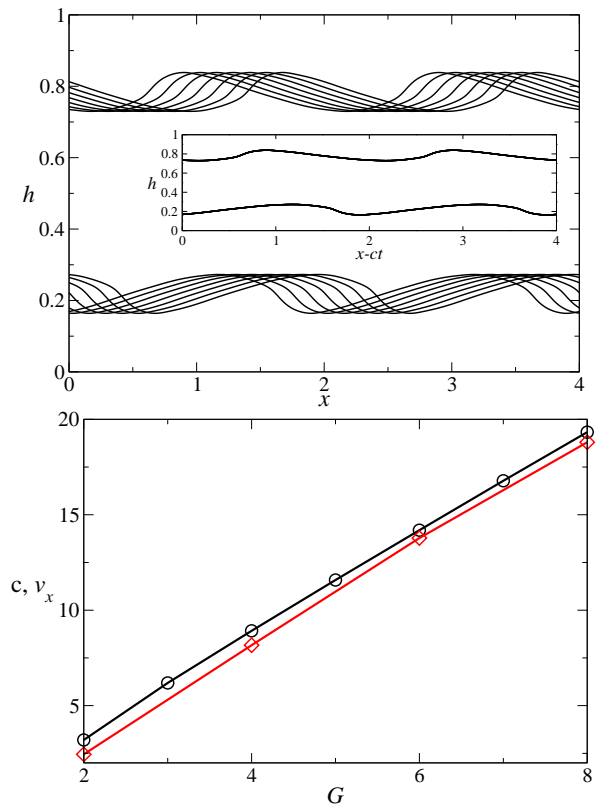

  \includegraphics[width=7.5cm]{./interface_G2.0_Lx4.0.eps}
  \includegraphics[width=7.8cm]{./speeds.eps}
  \caption{{\bf Top)} Snapshot of interface height $h(x)$ for $a=0.3$,
    $\eta=0.05$, $l=0.005$, $L_x=4.0$ and $G=2.0$ at six times equally
    spaced by $\Delta t=0.05265$.  $Dt=0.000025$, $N_x=200$,
    $N_y=400$. Inset: the same data plotted versus transformed
    coordinate $x-ct$ revealing simple convective motion with a
    constant wavespeed $c$.  \newline {\bf Bottom)} Diamonds:
    wavespeed $c$ versus pressure drop $G$, obtained by performing the
    transformation shown in Fig.~\ref{fig:interface}. $a=0.3$,
    $\eta=0.05$, $l=0.005$, $L_x=4.0$ for runs with $Dt=0.000025$, $N_x=200$,
    $N_y=400$. Circles: fluid velocity $v_x$ at interface, taken from
    the 1D banded profiles of Fig.~\ref{fig:base}.}
\label{fig:interface}
\end{figure}

% FILENAMES - TOP
% statePlots/state_G2.0_gdot0.0_eta0.05_a0.3_l0.005_Lx4.0_temperature1e-14_Dt0.000025_Nx200_Ny400_tmax100.0_faster_time60.000000001711_OPWxx.eps processed by interface_from_state.sh
% FILENAMES - BOTTOM
% statePlots/state_G2.0_gdot0.0_eta0.05_a0.3_l0.005_Lx4.0_temperature1e-14_Dt0.000025_Nx200_Ny400_tmax100.0_faster_time60.000000001711_OPWxx.eps
% statePlots/state_G4.0_gdot0.0_eta0.05_a0.3_l0.005_Lx4.0_temperature1e-14_Dt0.000025_Nx200_Ny400_tmax100.0_faster_time60.000000001711_OPWxx.eps
% statePlots/state_G6.0_gdot0.0_eta0.05_a0.3_l0.005_Lx4.0_temperature1e-14_Dt0.000025_Nx200_Ny400_tmax100.0_faster_time60.000000001711_OPWxx.eps
% statePlots/state_G4.0_gdot0.0_eta0.05_a0.3_l0.005_Lx6.0_temperature1e-14_Dt0.000025_Nx300_Ny400_tmax100.0_time60.000000001711_OPWxx.eps

In steady 1D PPF driven by a pressure drop $\partial_xP=-G$,
Eqn.~\ref{eqn:NS} dictates the shear stress to be linear across the
flow cell: $T_{xy}=-G(y-\tfrac{1}{2})$. Well away from any interfaces
(the almost vertical regions in Fig.~\ref{fig:base}, right) the value
of the shear rate $\gdot$ then follows as the lower (resp.\ upper)
solution branch of the same relation
$T_{xy}(\gdot)=\Sigma_{xy}(\gdot)+\eta\gdot$ discussed above for PCF,
in the regions where $|T_{xy}|<T_{\rm sel}$ (resp. $|T_{xy}|>T_{\rm
  sel}$). This leads to the steady flow states shown in
Fig.~\ref{fig:base}, which are shear banded for all values of the
applied pressure drop $G>2T_{\rm sel}$ shown. In bulk rheology, the
signature of the transition to shear banded flow is the pronounced
kink seen in the composite flow curve of Fig.~\ref{fig:flowCurve}
(right).

The 1D shear banded flows discussed in this section will form the basic
states and initial conditions for the stability studies of the rest of
the paper. In Sec.~\ref{sec:xy} we will consider 2D flow in the
$x$-$y$ plane, with periodic boundaries in the flow direction $x$, and
assuming translational invariance in $z$. In Sec.~\ref{sec:yz} we will
turn to 2D flow in the $y$-$z$ plane, with periodic boundaries in the
vorticity direction $z$, and assuming translational invariance in $x$.

\section{Flow, flow-gradient plane}
\label{sec:xy}

In this section we relax the assumption of translational invariance in
the flow direction $x$ and perform a two-dimensional study in the
flow/flow-gradient ($x$-$y$) plane. As before we consider flow between
closed walls at $y=\{0,L_y\}$, driven by a constant pressure drop
$\partial_xp=-G$. The boundaries in $x$ are taken to be periodic. For
simplicity, translational invariance is still assumed in the vorticity
direction $z$.

Our study comprises two parts. In Sec.~\ref{sec:xy_linear} we study
the linear stability of the 1D shear banded basic states of
Fig.~\ref{fig:base} with respect to perturbations that have
infinitesimal amplitude and a wavevector $q_x\hat{\vecv{x}}$ in the
flow direction. We will show these basic states to be linearly
unstable to such perturbations in most regimes. An initial condition
comprising shear banded flow with a flat interface is thereby
predicted to evolve towards a 2D state that has modulations along the
interface, via the growth of these perturbations.  In
Sec.~\ref{sec:xy_nonlinear} we study the model's full nonlinear
dynamics in this $x$-$y$ plane, giving results for the ultimate 2D
flow state that is attained once these modulations have grown to, and
(as we shall demonstrate) saturated at, a finite amplitude.

Before presenting our results we discuss briefly our numerical
method. In previous work the full nonlinear dynamics of the DJS model
in boundary driven PCF were simulated by SMF in the flow/flow-gradient
plane~\cite{fielding-prl-96--2006}, assuming translational invariance
in the vorticity direction. During the course of that study, the code
was carefully checked against an earlier calculation of the linear
stability of an interface between shear bands in
PCF~\cite{fielding-prl-95--2005,wilson-jnfm-138-181-2006}, providing a
stringent check that the nonlinear code was working correctly. A
straightforward two-line modification adapts that code to the pressure
driven PPF of interest here.

\subsection{Linear stability analysis}
\label{sec:xy_linear}

In each run of this modified code we start with an initial condition
comprising a 1D banded state of Sec.~\ref{sec:base}, corresponding to
shear banded flow with a flat interface between the bands. To this
state we add 2D perturbations of tiny amplitude.  By monitoring, in
this fully nonlinear code, the early-time growth of the Fourier
components $\exp(iq_xx)\exp(\omega t)$ of the perturbation, we can
extract the dispersion relation $\omega(q_x)$ that characterises the
linear instability of the 1D basic state. These numerical results for
this dispersion relation are shown by dashed lines in
Fig.~\ref{fig:dispersion} (top), for different values of the
interfacial thickness $l$.  As can be seen, at any fixed value of
$q_x$ the growth rate shows a linear dependence on $l$:
$\omega^*(l,q_x)=\omega^*(l=0,q_x)-a(q_x)l$ where the intercept
$\omega^*(l=0,q_x)$ plotted versus $q_x$ accordingly forms the
dispersion relation extrapolated to $l=0$, shown by the crosses in
Fig.~\ref{fig:dispersion} (top). The value $l=0$ cannot be accessed in
this full nonlinear code, because this limit is pathological in shear
banding fluids. However, analytical linear stability calculations can
nonetheless still be performed in this
limit~\cite{wilson-jnfm-138-181-2006}. The solid line in
Fig.~\ref{fig:dispersion} (top) accordingly shows the results of a
truly linear analytical stability calculation, performed by HJW in the
true limit $l\to 0$. As can be seen, this indeed agrees well with the
extrapolation of the numerical results to $l=0$.  Because the $l\neq
0$ and $l=0$ calculations were performed independently by the two
authors, and using different methods, this provides a good cross check
between both sets of results.

Because of the linear nature of this stability system, any eigenvector
can be separated into sinuous and varicose contributions, in which the
perturbation to the flow field across the channel is odd (snake-like)
and even (sausage-like) respectively. Each of these is an eigenvector
in its own right. In the $l=0$ calculations, this separation is
carried out explicitly. We find that the sinuous (snakelike) modes are
always more unstable than varicose ones. Because our interest is in
the most unstable mode, the $l=0$ results presented here are all for
sinuous perturbations. (In our runs of the full code for $l\not=0$, we
do not impose any such symmetry a priori. Instead, it emerges
naturally from the full system.)

The location of the peak in the dispersion relation is shown as a
function of pressure drop $G$ in Fig.~\ref{fig:dispersion} (bottom),
for the $l=0$ calculation. As can be seen, for values of $G$ well
within the banding regime the growth rate $\omega=O(1)$, comparable to
the linear viscoelastic relaxation time, and the wavelength
$\lambda_x=2\pi/q_x=O(2)$, comparable to twice the gap width. This is
consistent with the presence of secondary velocity
rolls~\cite{lerouge2008,fardin2009} of a wavelength comparable to the
gap width, and also consistent with the dominant wavelength seen in
our nonlinear results discussed in Fig.~\ref{fig:snapshots}
below. (Such velocity rolls will be shown explicitly in our
corresponding study of the flow-gradient/vorticity plane in
Fig.~\ref{fig:snapshots_z} below.)  As the applied shear rate
approaches the low shear branch from above, and the width of the high
shear bands accordingly tends to zero, the growth rate and wavelength
of the most unstable perturbations also tend to zero.

For a small range of pressure drops $1.20 < G < 1.52$, two peaks are
evident in the dispersion relation (not shown). For $G\le1.3$ the peak
with the larger value of $q_x$ is the more unstable, with crossover to
dominance of the lower $q_z$ peak for $G\ge1.32$. This causes a
``kink'' in the plot of $\omega^*$ against $G$ at $G\approx 1.3$, only
just discernible in Fig.~\ref{fig:dispersion}.

\subsection{Ultimate nonlinear dynamics}
\label{sec:xy_nonlinear}

At long times the instability described above saturates at the level
of finite amplitude undulations along the interface.  Representative
greyscale snapshots of the system's eventual state are shown in
Fig.~\ref{fig:snapshots}. As can seen in Fig.~\ref{fig:interface}
(top), the interfacial undulations convect along the positive $x$
direction with constant speed. This speed is plotted as a function of
pressure drop in Fig.~\ref{fig:interface} (bottom), and is comparable,
but not exactly equal, to the value of the fluid velocity at the
interface. As noted above, our $l\neq 0$ code does not impose any
sinuous/varicose symmetry. Indeed, the ultimate nonlinear attractor
can be seen in Fig.~\ref{fig:snapshots} to be dominated by a mode that
is neither fully sinuous not fully varicose.

\section{Flow-gradient, vorticity plane}
\label{sec:yz}

\begin{figure}[tb]
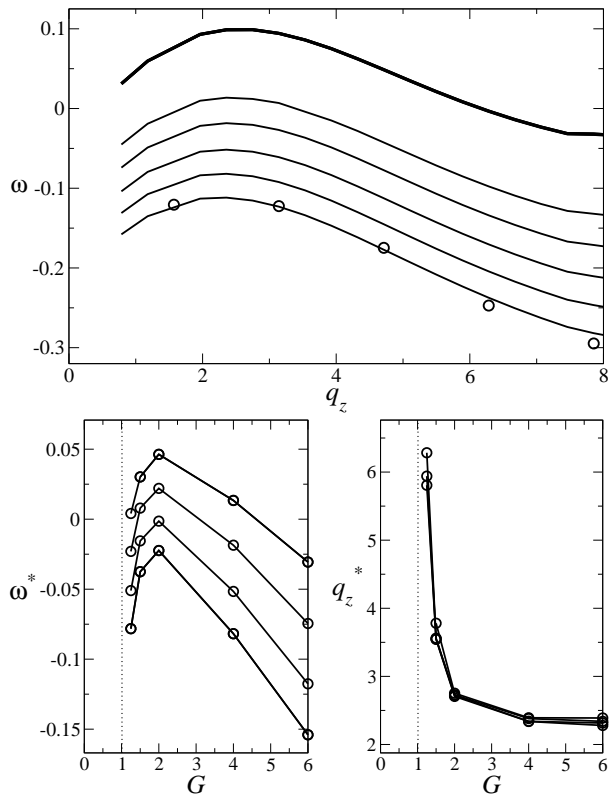

  \includegraphics[width=8cm]{./dispersion_z.eps}
  \includegraphics[width=8cm]{./peak_z.eps}
  \caption{{\bf Top)} Dispersion relation of growth rate $\omega$
    versus wavevector $q_z$ during the initial stage of 2D
    instability, starting with a 1D shear banded planar Poiseuille
    basic state. In each case $a=0.3$, $\eta=0.05$, $G=4.0$. Thin
    solid lines: calculated by linear stability analysis using the
    method of Ref.~\cite{fielding2007b}, with lines upward
    corresponding to $l=0.0035$, 0.003, 0.0025, 0.002, 0.0015.  Thick
    solid lines: extrapolation of the $l\neq 0$ results shown by each
    of the thin solid lines to $l=0$ using the scaling
    $\omega(q_x,l)=\omega(q_x,l=0)-a(q_x)l$. Indistinguishability of
    the thick lines demonstrates the validity of this scaling.
    Circles: extracted from early time dynamics of the full nonlinear
    code for $l=0.0035$.\newline {\bf Bottom)} Growth rate (left) and
    wavevector (right) at the peak of dispersion relation as a
    function of the pressure drop $G$ for $l=0.003$, 0.0025, 0.002,
    0.0015 (sets upwards in left hand figure).}
\label{fig:dispersion_z}
\end{figure}

%Linear dispersion relation at l=0.0035,..0.0015 and extrapolation to l=0.
%%% nonlinear results
% results/newWeirdInstability/vorticity_G4.0_gdot0.0_eta0.05_a0.3_l0.0035_Lz4.0_temperature1e-10_Dt0.05_Nz428_Ny457_tmax400.0_linear
%%% linear calculation
%interface/linearise/resultsPoiseuille/fixed0_q0.0_gdot0.0_eta0.05_a0.3_tauRlin5.0e-2_l0.0*_m0.0_k*_Nadapt100_G4.0_FPPF3.0, listing all eigenvalues, sorting, then throwing away nonsense ones near zero using the script processPoiseuile.sh
%%% extrapolation performed by scripts/extrapolateDispersion.sh

In this section we turn attention to the flow-gradient/vorticity
($y$-$z$) plane, now assuming translational invariance in the flow
direction $x$. As usual we assume closed walls in $y$, and (for
simplicity) periodic boundaries in $z$. We will return in the
conclusion to discuss briefly the implications of having focused only
on 2D studies in this paper.

\subsection{Linear stability analysis}
\label{sec:yz_linear}

We consider first the linear stability properties of the 1D basic
states of Fig.~\ref{fig:base} with respect to fluctuations with
wavevector in the vorticity direction, $\exp(i q_z z + \omega t)$. To
do so we perform a truly linear calculation, calculating the
eigenmodes of the stability matrix that is obtained by linearising the
full nonlinear equations about the basic state. This linearisation
problem was in fact studied previously in the flow-gradient/vorticity
plane in the context of planar Couette flow in
Ref.~\cite{fielding2007b}. The same linearised code can be used here,
simply inserting the new pressure driven basic state as an input to
the elements of the matrix.

The resulting dispersion relations $\omega(q_z)$ are shown in
Fig.~\ref{fig:dispersion_z} (top). As is evident, at each value of
$q_z$ the growth rate increases linearly with decreasing values of the
interfacial width $l$, allowing us to extrapolate to the case $l=0$
(thick line in Fig.~\ref{fig:dispersion_z}, top), as in the case of
the $x$-$y$ plane above. The location $\omega^*,k^*$ of the peak in
the dispersion relations is plotted as a function of the pressure drop
$G$ in Fig.~\ref{fig:dispersion_z}, bottom. As can be seen, for large
values of the interfacial width $l$ the instability is absent (all
$\omega^* < 0$). For intermediate values of $l$ there is a window in
$G$ between the onset of shear banding and the onset of this
interfacial instability. The size of this window decreases with
decreasing $l$, and we believe would extrapolate to zero as $l\to 0$,
such that all shear banded states are linearly unstable in the limit
of a thin interface.

These truly linear results were checked for consistency against the
linearised dynamics of a fully nonlinear code, which directly
simulates the full 2D dynamics of the model in the $yz$ plane. (The
main use of this code is to generate the results in the next
subsection.) In each run of this nonlinear code we used as an initial
condition the 1D basic state of Fig.~\ref{fig:base}, subject to small
2D fluctuations in the $y$-$z$ plane.  By monitoring the early-time
growth of the Fourier components $\exp(iq_zz)\exp(\omega t)$ we
extracted the dispersion relation $\omega(q_z)$ for the linear
(in)stability of the 1D basic state.  As shown by the circles in
Fig.~\ref{fig:dispersion_z}, good agreement is found with the true
linear stability calculation described in the previous paragraph. This
provides a stringent cross check between our linear stability
calculation and our nonlinear code.

\subsection{Ultimate nonlinear attractor}
\label{sec:yz_nonlinear}

\begin{figure}[tb]
  \includegraphics[width=8.0cm]{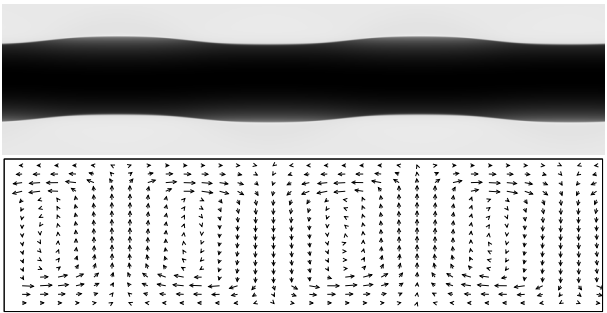}
  \includegraphics[width=8.0cm]{./flowMap.eps}
  \caption{Order parameter $\Sigma_{xx}(y,z)$ greyscale (top) and
    velocity map showing secondary flow in $y$-$z$ plane (bottom):
    snapshot at time $t=1895$ in steady state for $a=0.3$, $\eta=0.05$,
    $l=0.0015$ for $L_x=4.0$ and $G=2.0$ with $Dt=0.05$, $N_z=1000$,
    $N_y=1066$. }
% results/stateFiles/vorticity_G2.0_gdot0.0_eta0.05_a0.3_l0.0015_Lz4.0_temperature1e-10_Dt0.05_Nz1000_Ny1066_tmax2000.0_linear
\label{fig:snapshots_z}
\end{figure}

In previous work we simulated the full nonlinear dynamics of the DJS
model in 2D plane Couette flow in the flow-gradient/ vorticity
($y$-$z$) plane~\cite{fielding2007b}, assuming translational
invariance in the flow direction $x$. A straightforward two-line
modification allows this code to be adapted to the case of pressure
driven flow of interest here. As described in the previous section, we
performed a series of runs of this code starting with a 1D basic
state, subject to small 2D perturbations.  We consider in this section
the ultimate state that the system attains at long times.  In each
case, the instability was found to saturate in a steady state with
finite amplitude undulations along the interface
(Fig.~\ref{fig:snapshots_z}, top).  Such undulations have recently
been imaged experimentally in the pressure driven flow of wormlike
micelles in microchannels of high aspect
ratio~\cite{submitted}. Associated with these undulations is a
secondary flow comprising the velocity rolls shown in the second
subfigure (Fig.~\ref{fig:snapshots_z}, bottom).

%From any such pattern we can extract the amplitude of the interfacial
%undulations, which will serve as a convenient measure of the amplitude
%of the perturbation to 1D flow that exists in the final 2D
%pattern. This is plotted as a function of pressure drop in Fig.

%

\section{Summary and outlook}
\label{sec:conclusion}

In this work, we have studied the pressure driven planar Poiseuille
flow of a shear banding fluid sandwiched between infinite stationary
flat parallel plates. Using a combination of linear stability analysis
and direct numerical simulation, we have shown an initially flat
interface between shear bands to be unstable with respect to the
growth of undulations along it. The early time growth rate of these
undulations scales as the reciprocal stress relaxation time of the
fluid, and the corresponding wavelength is comparable to the
separation of the plates. At long times the undulations saturate in a
finite amplitude, cutoff by the nonlinear effects of shear.

Throughout our study we have neglected any in-flow and out-flow
effects at the start and end of the channel, assuming a well
established flow field that is free from end effects. Associated with
this is our assumption that the spatial limits relevant to this
established flow field correspond to the temporal limit $t\to \infty$
in each run of our code.  We have furthermore neglected the effect of
any lateral walls in the $z$ direction, taking from the outset the
limit $L_z/L_y\to\infty$. In practice one might expect lateral walls
to impose a further loss of translational symmetry in the $z$
direction, as explored in Ref.~\cite{submitted}. Perhaps most
importantly, in conducting separate two-dimensional studies in the
$x-y$ and $y-z$ planes, we have neglected the possibility that exists
in three spatial dimensions of nonlinear interactions between the
$q_x$ and $q_z$ modes. Indeed, although the $q_x$ modes have much
faster growth rates than the $q_z$ modes in the early time (linear)
growth regime, they are cut off more aggressively at nonlinear order
by the effects of shear.  This results in comparable amplitudes for
the $q_x$ and $q_z$ models in the ultimate attractors of our separate
2D studies. It therefore seems unlikely that 3D effects cannot be
neglected, as we shall explore in future. We mention finally that the
constitutive model used in the present work is highly oversimplified,
in particular in having a Newtonian high shear branch.

Recent experiments in microchannel flow used 1D velocimetry imaging to
reconstruct the local flow curves of a shear banding wormlike micellar
surfactant solution~\cite{masselon2008}. Surprisingly, these curves
were found to be non-universal, failing to collapse when plotted for
different applied pressure gradients on a single plot. Possible
explanations of this observation, to be explored in future work,
include: a larger ratio $l/L_y$ than considered here; coupling between
flow and concentration; and the effects of secondary 3D flows.

\section{Acknowledgements}

SMF thanks Tony Maggs for his hospitality during a research visit,
funded by the CNRS of France, to the Laboratoire de Physico-Chimie
Th\'{e}orique (PCT) at ESPCI in Paris, where this work was partly
carried out. SMF also thanks Philippe Nghe and Patrick Tabeling in the
Microfluidique, MEMS et Nanostructures group at ESPCI, and Armand
Ajdari of PCT for stimulating discussions. Financial support for SMF
is acknowledged from the UK's EPSRC in the form of an Advanced
Research Fellowship, grant reference EP/E5336X/1.

 %%%%%%%%%%%%%%%%%%%%%%%%%%%%%%%%%%%%%%%%%%%%%%%%%%%%%%%%%%%%%%%%%%%%%%%%%%%%%

%\bibliographystyle{prsty}
%\bibliography{ackerson,actin,articles,banding,barham,Berret,berrportdecruppe,berthier,books,bray,callaghan,cates,chandcits,cook,crystal,crystal_theory,Decruppe,dhont,dnatheory,elasticTurbulence,fielding,fischer,Fischer,flowcryst,fredrickson,gelbart,goveas,graham,Groisman,head,hebraud,helfrich,HinchRallison,hsiao,Kadoma,larson,LCtheory,leal,lerougeDecruppe,lifshitz,line_tension,maffettone,malkus,master,mccoy,membs,Mexican,new,noirez,notes,olmsted,onions,otherRelated,phanThien,phd1,phd,pine,PineHu,pomeau,poon,pratt,psolutions,ramaswamy,recent,rheochaos,rheofolks,ryan,salmon,SalmonManneville,savedrecs.txt,schoot,semenov,sgrband,shaqfeh,sood,sriram,stein,sureshkumar,vansaarloos,vorticity,Wang,weitz,wilson,worms2,worms3,worms,yuan,zubarev,worms3}

%%%%%%%%%%%%%%%%%%%%%%%%%%%%%%%%%%%%%%%%%%%%%%%%%%%%%%%%%%%%%%%%%%%%%%%%%%%%%
\end{document}